\documentclass[pra,oneocolumn,showpacs,groupedaddress,superscriptaddress,nofootinbib,floatfix,preprintnumbers,longbibliography]{revtex4-2}
\usepackage{amssymb,amsmath,graphicx,color}\usepackage{bm}
\usepackage[tight]{subfigure}
\usepackage[export]{adjustbox}
\usepackage{braket}
\usepackage[colorlinks=true,citecolor=blue,linkcolor=red]{hyperref}
\usepackage{multirow}
\usepackage{rotating,booktabs}
\usepackage[verbose]{placeins}
\usepackage{physics}
\usepackage{dsfont}

\begin{document}

\title{Quantum Optics as Applied Quantum Electrodynamics is back in town}

\author{Philipp Stammer}

\affiliation{ICFO-Institut de Ciencies Fotoniques, The Barcelona Institute of Science and Technology, Castelldefels (Barcelona) 08860, Spain}

\author{Maciej Lewenstein}
\email{maciej.lewenstein@icfo.es}
\affiliation{ICFO-Institut de Ciencies Fotoniques, The Barcelona Institute of Science and Technology, Castelldefels (Barcelona) 08860, Spain}

\affiliation{ICREA, Pg. Lluis
Companys 23, 08010 Barcelona, Spain}

\begin{abstract}
We start this short note by remembering the beginnings of the Warsaw School of Quantum Optics, evidently stimulated  by Iwo Bia\l ynicki-Birula at the Warsaw University, and then Centre for Theoretical Physics of Polish Academy of Sciences, and Adam Kujawski and Zofia Bia\l ynicka-Birula at the Institute of Physics of Polish Academy of Sciences. In  the theoretical approaches of the Warsaw School Quantum Field Theory was always present, and Quantum Optics was considered to be Applied Quantum Electrodynamics (QED). All of us who grew up in this fantastic community have carried and are still carrying the gospel to others. In particular, now QED began her run on the red carpet of Super Instense Laser Matter Interactions, Attosecond-physics, and Ultrafast Laser Physics, in general. We will elaborate on the recent progress in this direction, and on the open questions towards future investigations. This paper celebrates the 90th birthday of Prof. Iwo Bia\l ynicki-Birula, our QED guru! 
\end{abstract}

\date{\today}

\maketitle


\section{Introduction}

\subsection{Memories}

On the occasion like this, it is appropriate to start the paper with some personal memories, in this case by M. Lewenstein: Me and one of my best friends, Marek Ku\'s were supposed to do our Diplomas at the Department of Physics of Warsaw University in the academic year 1978-1979. Like many other top theory students our preference was Katedra Metod Matematycznych Fizyki (KMMF), led by Prof. Krzysztof Maurin. I even had a favorite supervisor: Krzysztof Gaw\c edzki. When I asked him about the possibility he told me literally: "Panie Ma\'cku, Quantum Field Theory is difficult, and Renormalization Group even harder", and he left Poland starting his Odyssey via Harvard, Princeton, IHES, and ENS Lyon.
Still, we wanted to go to KMMF, but the Dean of the Department, Prof. Jerzy Pniewski issued a rule that there would be no Diplomas in KMMF this year. We had to look for something comparably challenging, and we chose Zak\l ad Teorii Pola i Fizyki Statystycznej of Prof. Iwo Bia\l ynicki/Birula, the author of the seminal handbook of Quantum Electrodynamics \cite{bialynicki2013quantum}. It was indeed a Mekka of the Warsaw Statistical Physics with Jarek Piasecki, \L ukasz Turski and Bogdan Cichocki, but we were interested in Quantum Field Theory (QFT). And then came two younger and very convincing guys, Kazik Rz\c a\.zewski and Krzysztof Wódkiewicz, who saiud: let us do Quantum Optics (QO), which is Applied QED. And we both got seduced. 

Indeed, training of QO in Warsaw was heavily biased  toward QFT. Master equation approaches were not "allowed", one was using full Hamiltonian and Heisenberg equations. This has taught us very early that there are no Markov processes in Nature: everything must have long time tail corrections and more...

There is another twist to this story related to Stong Laser Field physics. On the desk of Kazik Rz\c a\.zewski, I found a preprint of Luiz Davidovich that Kazik got when they shared the same bureau at ITCP with Luis. I got absolute fascinated by the Keldysh's theory of tunnel ionization,  and decided to work on it. In the beginning of 1970, Pierre Agostini in  Saclay published first result on, the so called, Above Threshold Ionisation. Zofia Bia\l ynicka-Birula published a seminal paper \cite{zofia} on the subject in 1984. This is the moment, when I decided to join operation. 

The situation of the Super Intense Laser Matter Physics is well described in the sub-section below. We clearly face the situation when QED is on the run again. 
This paper is based on the thesis proposal of Philipp Stammer, a PhD student at ICFO. So, the plan is to present the motivation to bring Quantum Optics as Applied Quantum Electrodynamics back to town, and then various future projects, all related to QED of Strong Laser Fields physics, so to the clear heritage of Iwo Bia\l ynicki-Birula.

\subsection{\label{sec:QO_meets}Quantum Optics meets Strong Laser Field Physics}

For decades the interaction of intense and short laser pulses with matter has been described successfully with semi-classical methods, in which the quantum nature of the electromagnetic field was not taken into account. The characteristics of the observed features in the spectra for the processes of high harmonic generation (HHG) \cite{ferray1988multiple, lewenstein1994theory} or above threshold ionization (ATI) \cite{agostini1979free, lewenstein1995rings} were well reproduced within the semi-classical picture. 
Furthermore, the semi-classical approach for the process of HHG (or even fully classical \cite{corkum1993plasma}) provides a powerful picture by means of the so-called 3-step model to gain intuition about the electron dynamics. There, (i) an electron tunnel ionizes into the continuum through the barrier formed by the Coulomb potential of the core and the electric field (via dipole coupling), then (ii) the freed electron is driven in the presence of the electric field and can (iii) eventually recombine to the core by emitting the gained energy in terms of radiation. 
This description has lead to fruitful analysis in terms of quantum trajectories \cite{salieres2001feynman, ivanov2005anatomy, smirnova2013multielectron} within the strong field approximation \cite{amini2019symphony}. The progress of the strong field and attosecond physics based on the semi-classical description as immense, but neglecting the quantum properties of the field did not allow to use a language for posing specific questions on the dynamics. \\

However, including the quantum electrodynamical characteristics of the field can lead to new observations in the radiation field inaccessible from the classical perspective, and further allows to ask question unamenable before, for instance to obtain insights about the quantum state of the field. In fact, recent theoretical and experimental advances have indicated that intense laser matter interaction can exhibit non-classical features. In particular, quantum optical approaches of the process of high-order harmonic generation asked for the quantum state of the harmonic field modes \cite{gorlach2020quantum, lewenstein2021generation}, and studied the back-action on the fundamental driving field \cite{lewenstein2021generation, rivera2022strong}. Furthermore, the experimental advances in combining strong field physics with methods known from quantum optics \cite{gonoskov2016quantum, tsatrafyllis2017high}, allowed to conceive new experiments in which non-classical states of light can be generated from the HHG process \cite{lewenstein2021generation, rivera2022strong, stammer2022high}.
This progress has then triggered subsequent analysis for quantum state engineering of light using intense laser matter interaction \cite{stammer2022theory, stammer2022quantum, rivera2022light}. Nevertheless, and despite using Hilbert space constructs for the electromagnetic field, the investigation has yet not revealed inherent quantum signatures in the emitted radiation from the HHG process itself.  

Besides these achievements in the quantum optical description of intense laser driven processes, the full quantum optical properties of the emitted radiation in the process of high harmonic generation has yet not been revealed. The radiation is obtained from classical dipole antenna like sources, and thus exhibit the same characteristics as classical coherent radiation sources. Furthermore, the quantum state of the electromagnetic field is given in terms of product coherent states, which are classical states. Those features originate from the neglected dipole moment correlations in the current theory \cite{sundaram1990high, lewenstein2021generation, stammer2022theory, stammer2022quantum}, which, if taken into account, would eventually lead to non-classical contributions in the properties of the emitted harmonic radiation. 
Thus, further investigation towards accessing this information, with potential hidden and interesting properties seems promising for a more detailed understanding of the HHG process and for potential applications in optical technologies.  
Nevertheless, by introducing conditioning measurements on the field after the HHG process leads to the generation of non-classical field states by means of optical Schrödinger cat states with high photon numbers \cite{rivera2022strong, stammer2022high, stammer2022theory, stammer2022quantum}. This suggest potential applicability of this methods in modern optical quantum technologies, and could provide a new photonic platform for information processing \cite{lewenstein2022attosecond, bhattacharya2023strong}. 
In particular since quantum information processing often requires entangled or superposition states as a resource, there is a clear need to generate such states. \\

The next section provides an introduction to the current quantum optical formulation of the process of high harmonic generation. 
This serves to define the stage for introducing current open questions the new formalism caused. This will then allow to propose further investigation in this direction. In particular it highlights the assumptions and approximations used, which are then questioned and analyzed in the proposed future analysis.

\subsection{\label{sec:intro}Quantum optical high harmonic generation}

In the process of high harmonic generation, coherent radiation of higher order harmonics of the driving laser frequency is generated \cite{lewenstein1994theory, salieres1997study}. The transfer of coherence, and energy, from the intense laser source to the harmonic field modes (initially in the vacuum) is achieved by a highly nonlinear interaction of the driving field with the HHG medium, in which the electron is used as an intermediary between the optical modes. Until recently this was mainly described in semiclassical terms, in which only the electronic degrees of freedom are quantized \cite{lewenstein1994theory}, although there have been early approaches to introduce a fully quantized description of the HHG process \cite{sundaram1990high, eberly1992spectrum, diestler2008harmonic}.
However, recent advances in the quantum optical analysis of HHG has established a new direction in the investigation of strong field physics. This allows to study the quantum mechanical properties of the harmonic radiation, or to take into account the backaction on the driving field \cite{kominis2014quantum, gonoskov2016quantum, tsatrafyllis2017high, gorlach2020quantum, lewenstein2021generation, rivera2022strong, stammer2022high, stammer2022quantum, stammer2022theory, rivera2022light}. In particular, it was shown that conditioning procedures on processes induced by intense laser-matter interaction can lead to the generation of high-photon number controllable non-classical field states in a broad spectral range \cite{lewenstein2021generation, rivera2022strong, stammer2022high, stammer2022quantum, stammer2022theory}. \\

What now follows is a brief introduction to the quantum optical description of the process of HHG. We will consider discrete field modes for the sake of simplicity, and would like to refer the reader to the full quantum-electrodynamical description including a continuum of field modes given in \cite{stammer2022quantum}. 
To describe the process of HHG in the single-atom picture (see \cite{sundaram1990high} in which case this is legitimate) we assume that a single active electron is initially in the ground state $\ket{g}$, and is driven by a strong laser field which is described by a coherent state $\ket{\alpha}$ in the fundamental driving mode. The harmonic field modes $q \in \{2,..., N\}$ are initially in the vacuum $\ket{\{ 0_q \}} = \otimes_{q \ge 2} \ket{0_q}$.
The interaction Hamiltonian describing the process in the length-gauge, and within the dipole approximation, is given by 
\begin{align}
    H_I(t) = - \vb{d}(t) \cdot \vb{E}_Q(t),    
\end{align}
where the electric field operator $\vb{E}_Q(t) = - i g \sum_{q=1}^N \sqrt{q} \left( b_q^\dagger e^{i q \omega t} - b_q e^{- i q \omega t} \right)$ is coupled to the time-dependent dipole moment operator
\begin{align}
    \vb{d}(t) = U_{sc}^\dagger (t,t_0) \vb{d} U_{sc}(t,t_0).    
\end{align}
The dipole moment is in the interaction picture of the semi-classical frame $U_{sc} (t,t_0) = \mathcal{T} \exp [- i \int_{t_0}^t d\tau H_{sc}(\tau)]$, with respect to the Hamiltonian of the electron 
\begin{align}
    H_{sc}(t) = H_A - \vb{d} \cdot \vb{E}_{cl}(t).    
\end{align}
This semi-classical Hamiltonian is the same as traditionally considered in semi-classical HHG theory \cite{lewenstein1994theory}, where $H_A = \vb{p}^2/{2} + V(\vb{r})$ is the pure electronic Hamiltonian, and 
\begin{align}
    \vb{E}_{cl}(t) = \Tr[\vb{E}_Q(t) \dyad{\alpha}] = i g (\alpha e^{- i \omega t} - \alpha^* e^{i \omega t}),    
\end{align}
is the classical part of the driving laser field. A detailed derivation of the interaction Hamiltonian $H_I(t)$ can be found in \cite{stammer2022quantum}. 
It now remains to solve the time-dependent Schrödinger equation (TDSE) for the dynamics of the total system of electron and field. Since we are interested in the quantum optical dynamics of the field, and in particular on the process of HHG, we consider the field evolution conditioned on the electronic ground state (this is because the electron returns to the ground state in the HHG process). We thus project the TDSE on $\ket{g}$, and it remains to solve 
\begin{align}
    i \partial_t \ket{\Phi(t)} = - \bra{g} \vb{d}(t) \cdot \vb{E}_Q(t) \ket{\Psi(t)},
\end{align}
where $\ket{\Phi(t)} = \bra{g} \ket{\Psi(t)}$ with the state of the total system $\ket{\Psi(t)}$. 
Taking into account that the electron is initially in the ground state, it is equivalent to solve for the operator
\begin{align}
\label{eq:kraus_exact}
    K_{HHG} = \bra{g} \mathcal{T} \exp \left[ i \int_{t_0}^t dt^\prime \vb{d}(t^\prime) \cdot \vb{E}_Q(t^\prime) \right] \ket{g},
\end{align}
which solely acts on the initial field state $\ket{\Phi_i} = \ket{\alpha} \ket{\{ 0_q \}}$. This can be solved exactly when neglecting correlations in the dipole moment of the electron \cite{sundaram1990high, stammer2022theory}, such that we can write 
\begin{align}
\label{eq:kraus_approx}
    K_{HHG} \approx  \mathcal{T} \exp \left[ i \int_{t_0}^t dt^\prime \bra{g}\vb{d}(t^\prime)\ket{g} \cdot \vb{E}_Q(t^\prime) \right] = \prod_{q=1}^N e^{i \varphi_q} D(\chi_q),
\end{align}
where the shift in each mode is given by the respective Fourier component of the time-dependent dipole moment expectation value
\begin{align}
    \chi_q = - i g \int_{t_0}^t dt^\prime \expval{\vb{d}(t^\prime)} e^{i q \omega t^\prime}.
\end{align}
Thus, the solution to \eqref{eq:kraus_approx} is given by a displacement operation acting on the field modes 
\begin{align}
\label{eq:HHG_final}
    \ket{\Phi} = K_{HHG} \ket{\Phi_i} = K_{HHG} \ket{\alpha}\otimes_{q\ge 2} \ket{0_q} = \ket{\alpha + \chi_1} \otimes_{q \ge 2} \ket{\chi_q}.
\end{align}
That the harmonic modes are described by coherent states is due to the fact that the source for the coherent radiation is related to the electron dipole moment expectation value $\expval{\vb{d}(t)} = \bra{g} \vb{d}(t) \ket{g}$, which acts as a classical charge current coupled to the field operator. It does thus only represent the coherent contribution to the harmonic radiation field, and no genuine quantum signature is found. 
Furthermore, the fact that the final state is a product coherent state over all modes is a consequence of the approximation of neglecting the dipole moment correlations. Otherwise, if going beyond the linear order in $\vb{E}_Q(t)$, the field operators for different modes would mix when evaluating the exact propagator in \eqref{eq:kraus_exact} (see section \ref{sec:squeeze}). Nevertheless, a phenomenological approach to take into account the entanglement between the field modes was performed by the authors in \cite{stammer2022high, stammer2022theory}. 

\begin{figure}
    \centering
    \includegraphics[width=1.0\textwidth]{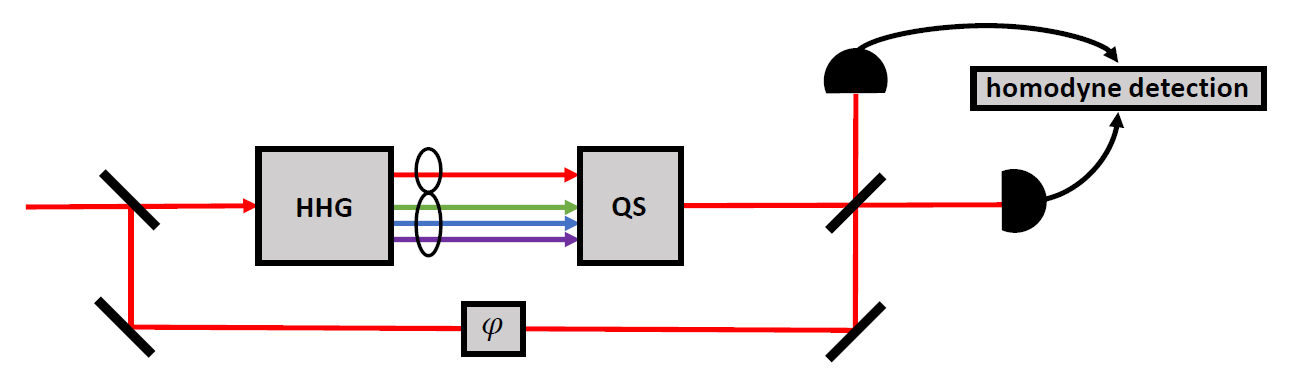}
    \caption{Schematic illustration of the HHG conditioning experiment performed to generate optical cat states with controllable quantum features. An intense laser field drives the process of HHG, in which an entangled state of the fundamental mode and all harmonics is generated. A conditioning measurement on the harmonic field modes in the quantum spectrometer (QS) leads to a coherent state superposition in the driving field of the form \eqref{eq:cat}, and is measurement with a homodyne detection scheme after overlapping with a local oscillator of varying phase delay $\varphi$. The reconstructed Wigner functions of the homodyne measurement are shown in \ref{fig:Wigner}.}
    \label{fig:setup}
\end{figure}

\begin{figure}
    \centering
    \includegraphics[width=0.4\textwidth]{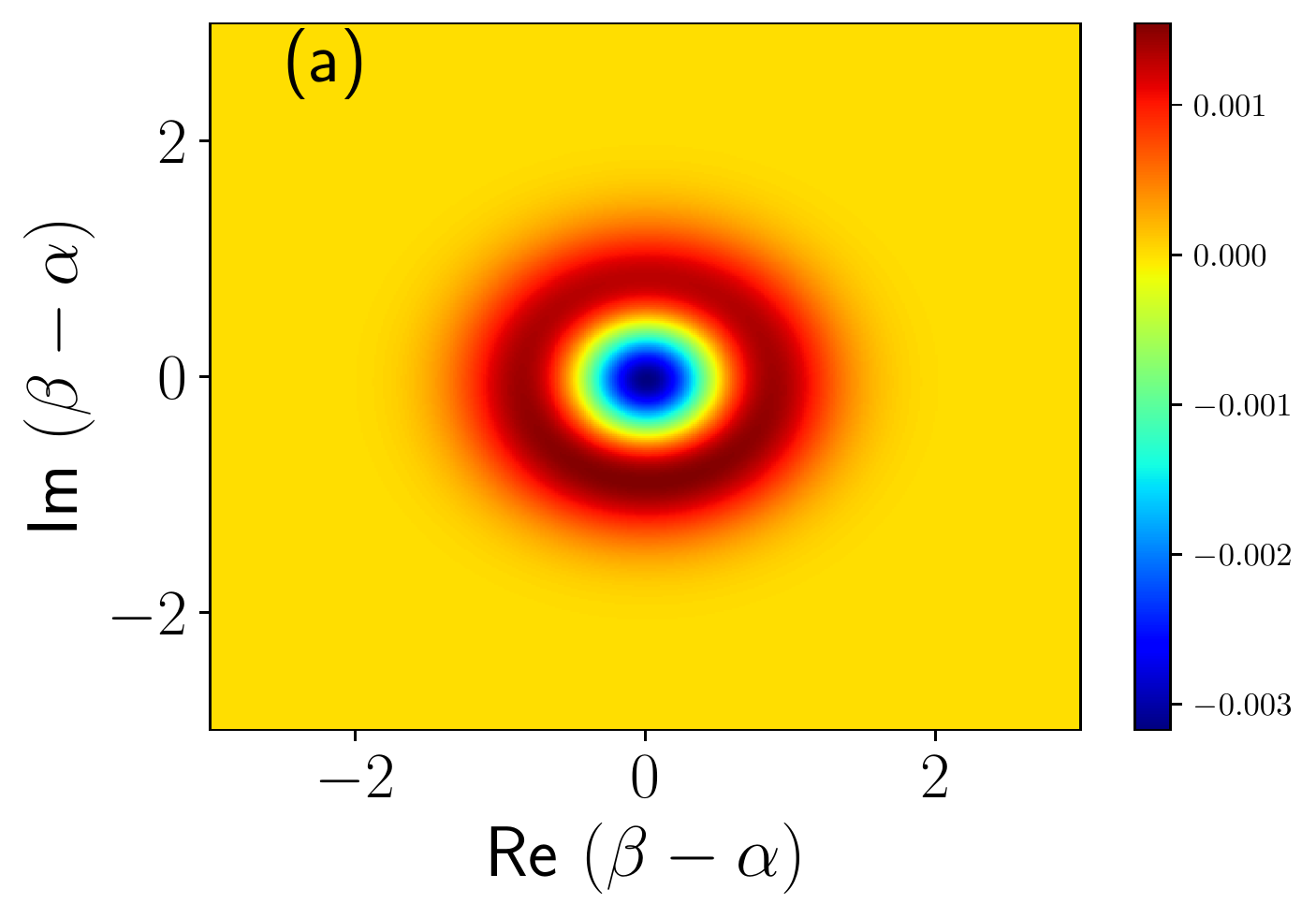}
    \includegraphics[width=0.4\textwidth]{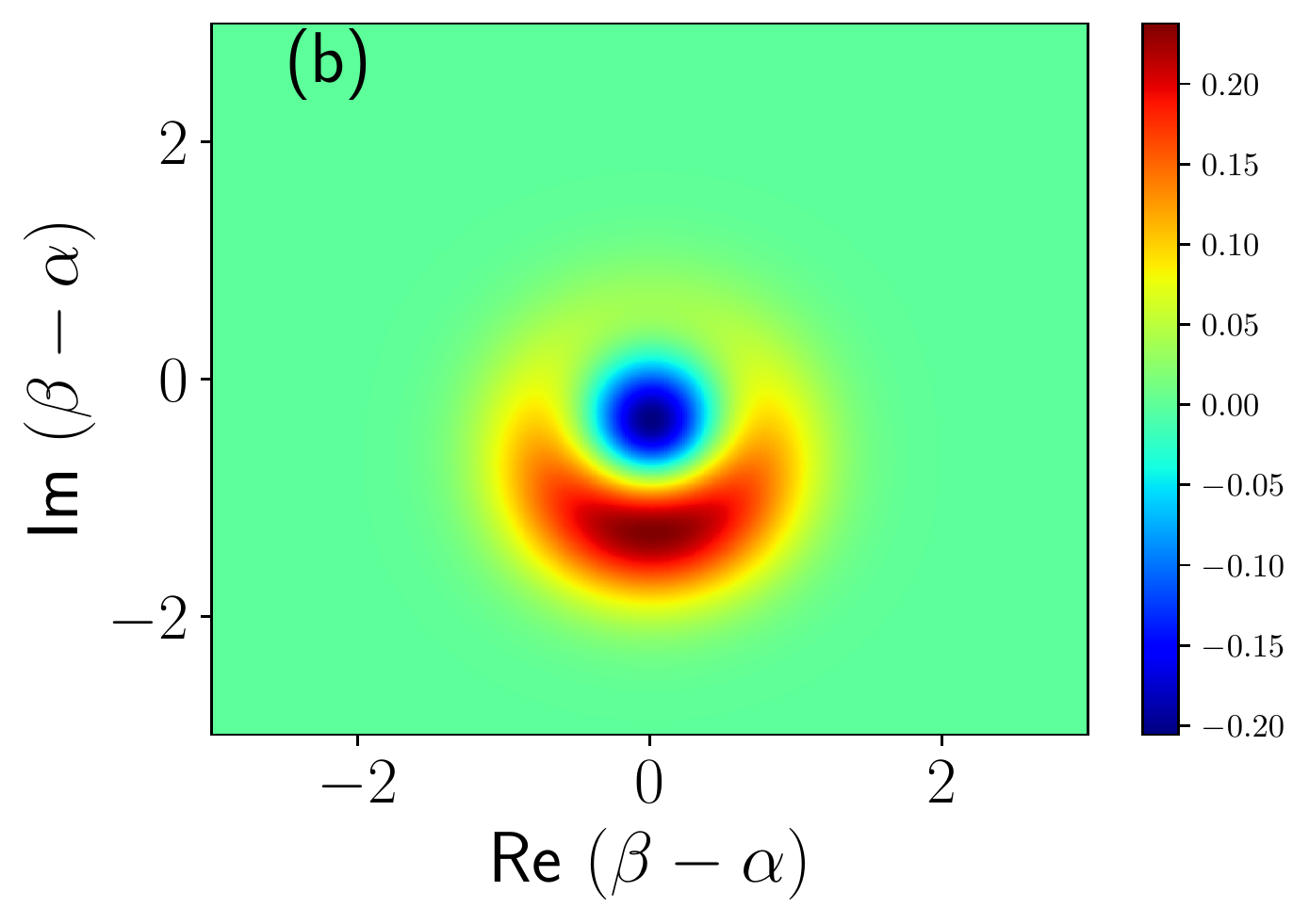}
    \caption{Wigner function of the coherent state superposition in \eqref{eq:cat} for different displacement of (a) $\chi_1 = 0.1$ (b) $\chi_1 = 1.0$, which shows features of an optical "kitten"-state and a "cat"-state, respectively. }
    \label{fig:Wigner}
\end{figure}

However, we can employ conditioning schemes on certain field modes which allows for quantum state engineering of light with non-classical properties \cite{stammer2022theory, stammer2022quantum}.
In particular, it was shown experimentally that a conditioning procedure on the process of HHG can lead to coherent state superposition states (CSS) in the driving laser mode (in the IR regime) in close analogy to optical cat states \cite{lewenstein2021generation, rivera2022strong}. The experimental configuration is schematically shown in Fig. \ref{fig:setup}, in which the the conditioning on HHG is carried out, and a homodyne detection measurement of the fundamental driving field is performed \cite{lewenstein2021generation, stammer2022quantum}.
To formally describe the generation of those optical CSS via a conditioning operation on the HHG state $\ket{\Phi} = \ket{\alpha + \chi_1} \otimes_{q\ge2} \ket{\chi_q}$ from \eqref{eq:HHG_final}, M. Lewenstein recognized that it can be obtained through the projection onto $P = \mathds{1}  - \dyad{\alpha}$. This projector was phenomenologically introduced in \cite{lewenstein2021generation}, and lead to the CSS state 
\begin{align}
\label{eq:cat}
    \ket{\psi} = \ket{\alpha + \chi_1} - \bra{\alpha}\ket{\alpha + \chi_1} \ket{\alpha}.
\end{align}
It was then shown by P. Stammer in \cite{stammer2022high, stammer2022theory} how this projector follows from a projective measurement on the harmonic field modes when further taking into account the correlations between the field modes, and also derived the actual measurement operation $M^\chi_\alpha = \mathds{1} - e^{- \sum_{q\ge2} \abs{\chi_q}^2} \dyad{\alpha}$, which converges to the projector $M_\alpha^\chi \simeq P = \mathds{1}  - \dyad{\alpha}$ since $\sum_{q\ge 2} \abs{\chi_q}^2$ is on the order $\mathcal{O}(1/N)$. The completeness relation of the associated positive operator-valued measure for the measurement operator was shown in \cite{stammer2022theory} within the framework of the quantum theory of measurement.
To reconstruct the quantum state of the coherent state superposition in \eqref{eq:cat} a homodyne detection measurement is performed (see Fig. \ref{fig:setup}), and the Wigner function of the state is reconstructed. The Wigner function corresponding to the CSS in \eqref{eq:cat} is shown in Fig. \ref{fig:Wigner} for two different values of the displacement $\chi_1$. The possibility of experimentally varying the displacement $\chi_1$, for instance by changing the gas density in the HHG interaction region, allows to change the CSS from an optical "kitten"-state for small displacement (displaced first Fock state) to an optical "cat"-state for larger displacement, as shown in Fig. \ref{fig:Wigner} (a) and (b), respectively. This allows to have control over the non-classical properties of the generated CSS in order to generate high-photon number optical cat states from the infrared to the extreme ultraviolet regime \cite{lewenstein2021generation, stammer2022high}. We note, that the displacement $\chi_1$ can not be arbitrarily large, since it would destroy the superposition in \eqref{eq:cat} due to the pre-factor in the second term which is given by the overlap of the two states in the superposition. However, since $\alpha$ is the initial amplitude of the coherent state, this value has very high photon number, and thus the optical cat and kitten states can life far away in phase space while the two states in the superposition are not distinguishable.

\section{\label{sec:questions}Open questions about Quantum Optics of High Harmonic Generation}

In the previous section we have outlined the current state of the art of our efforts to have a quantum optical description of the process of high harmonic generation. 
However, there we have made assumptions about the experimental boundary conditions, and performed approximations by neglecting particular contributions. Those need to be tested. Furthermore, the quantum optical description of the light-matter interaction has not yet revealed any genuinely quantum mechanical feature in the HHG emission process itself. It turned out, that the state of the harmonic field modes $\{ q\}$ are described by product coherent states $\ket{\chi_q}$ - which are purely classical. Non-classical signatures, by means of the optical cat state, emerged through the conditioning process. 
However, we believe that the emitted radiation in the process of HHG contains non-classical signatures once the incoherent contribution from the dipole moment correlations are taken into account, and furthermore, that the field state will be entangled. 

In the following we will outline some open questions in the description of the process of high order harmonic generation from a quantum optical point of view, and provide a motivation why this should be a matter of interest for future investigations.

\subsection{On the role of the optical phase in high harmonic generation}

To describe the experimental conditions of the HHG experiment, we have assumed that the radiation field which drives the process can be described by a single-mode coherent state $\ket{\alpha}$. This would imply that the source emits continuous coherent laser light in a single mode with a well defined phase (coherent in the sense of having non-vanishing off-diagonal density matrix elements in the photon number basis). However, standard HHG experiments are performed by using a pulsed source of radiation. 
On the one hand this would automatically require a multi-mode description in the frequency domain due to the finite duration of the pulses (they are not just finite, but rather super short in the regime of femtoseconds). And thus we extended the theory to a continuum of modes given in Ref. \cite{stammer2022quantum}.
Furthermore, assuming a pure coherent state description implies that the field has a well defined phase, and would thus require a phase-stabilized laser system, such that the carrier-wave and the envelope of the pulse have a fixed phase relation from shot to shot (CEP-stabilization \cite{krausz2009attosecond}). Otherwise, for non phase-stabilized driving lasers, where the phase varies from shot to shot, one has to average over all possible phases, and take into account a proper mixed initial state 
\begin{equation}
\label{eq:rho_mixed}
    \rho_{|\alpha|} = \frac{1}{2 \pi} \int_0^{2\pi} d \varphi \dyad{\alpha e^{i \varphi}} = e^{- \abs{\alpha}^2}\sum_n \frac{\abs{\alpha}^{2n}}{n!} \dyad{n}  .  
\end{equation}
In particular the experiments in Refs. \cite{lewenstein2021generation, rivera2022strong}, which uses the process of HHG to generate optical cat-states do not use CEP-stable driving fields. 
To analyze the process of HHG, and the conditioning experiment introduced in \cite{lewenstein2021generation}, without the assumption of having a pure coherent initial state within the current quantum optical description there arise formal difficulties and interpretational inconsistencies with the well accepted picture of the HHG process. 

The difficulty arising in the formal analysis is that the semi-classical frame from the interaction picture of the Hamiltonian $H_I(t)$ (see section \ref{sec:intro}) is not well defined for mixed initial states. Within a fixed semi-classical frame, which is defined via the unitary transformation $D(\alpha)$, we have seen that HHG effectively leads to a shift in the field modes, i.e. $\rho_0 \to K_{HHG} \rho_0 K_{HHG}^\dagger$ (see Eq.\eqref{eq:HHG_final}).
However, for the mixed state $\rho_{\abs{\alpha}}$ there is no well defined semi-classical frame defined from a unique displacement operation $D(\alpha)$.
This can also be seen from the fact that the classical part of the driving field vanishes
\begin{align}
    \vb{E}_{cl}(t) = \expval{\vb{E}_Q(t)} = \Tr[\vb{E}_Q(t) \rho_{\abs{\alpha}}] = 0,    
\end{align}
which implies that there is a vanishing mean electric field amplitude. Hence, this conflicts with the traditionally used powerful picture of HHG in terms of the 3-step model introduced in section \ref{sec:QO_meets}. In this picture the presence of a non-vanishing electric field amplitude is crucial for describing the tunnel ionization process and the electron dynamics in the continuum driven by the field. 
The underlying physical property, for the fact that the semi-classical frame is only uniquely defined for a pure coherent initial state $\ket{\alpha}$, is the phase of the field. A coherent state has a well defined phase, which implies that the semi-classical frame exists via
\begin{align}
    \vb{E}_{cl}(t) = \expval{\vb{E}_Q(t)} = \Tr[\vb{E}_Q(t) \dyad{\alpha}] = \bra{\alpha} \vb{E}_Q(t) \ket{\alpha} \propto \sin(\omega t)    
\end{align}
and the classical picture of an electric field driving the electron process holds.
However, it is now natural to ask, if the process of high harmonic generation requires non-vanishing field amplitudes as suggested by the 3-step model, and if harmonics can be generated from driving fields without optical coherence such as the phase randomized state in \eqref{eq:rho_mixed}, which is diagonal in the photon number basis. Such a state with vanishing off-diagonal density matrix elements in the photon number basis does not exhibit optical coherence, and we thus ask if optical coherence in the driving field is a necessary requirement to generate high-order harmonics.
For instance, the electric field expectation value of the mixed state \eqref{eq:rho_mixed} vanishes $\expval{\vb{E}_Q} = \Tr[\vb{E}_Q \rho_{|\alpha|}] = \vb{E}_{cl} = 0$, due to the totally arbitrary phase and thus there is no well defined semi-classical frame.  
This ultimately leads to the question if processes driven by sufficiently large photon number states $\ket{n}$, which have a completely random phase due to the well defined photon number, allows for the generation of high-order harmonics. Or, even more general, if incoherent radiation can be used to drive the parametric process of HHG as recently observed for spontaneous parametric down-conversion in \cite{li2022experimental}. 

In many optical experiments the presence of optical coherence is not required to explain the measurement results, and the question of the requirement of optical coherence was first posed in \cite{molmer1997optical}. 
It is thus natural to ask if the process of HHG requires optical coherence (in the sense of a non-diagonal density matrix in the photon number basis), or if an optical field with a vanishing mean electric field amplitude is sufficient to drive the HHG process? If this is not the case, and we can generate high-order harmonics with incoherent light, how do the harmonic radiation properties differ? And furthermore, how can the powerful picture of the 3-step model be understood for driving fields with vanishing mean field amplitude?
Those question suggest that there is a need for further theoretical investigation about the role of the optical phase in the HHG process, and further if the conditioning experiment in \cite{lewenstein2021generation} is sensitive to the phase of the field or not. 
From an experimental perspective, we are eager to observe the reconstruction of the Wigner function for CEP stabilized driving laser fields. 
From the theoretical point of view, the first question necessary to answer in order to describe the experimental boundary conditions: \textit{What is the quantum state of an ultrashort few-cycle (CEP stable) laser pulse?}
One way to approach this question could be by following the arguments similar to \cite{van2001quantum, van2001quantum2} or \cite{bartlett2006dialogue}, just for pulses of radiation with and without CEP-stabilization.

\subsection{Theory of quantum optical coherence of high harmonic generation}

In the derivation of the field state after the process of HHG we have thus far always neglected the correlations in the dipole moment of the electron, i.e. approximating \eqref{eq:kraus_exact} with \eqref{eq:kraus_approx}. Consequently, we only considered a classical charge (by virtue of the dipole moment expectation value) coupled to the field operator. Therefore, we have only considered the coherent contribution to the harmonic radiation field. This has the advantage of being exactly solvable. However, as commonly known \cite{carmichael1999statistical} the incoherent contribution of the emitted radiation can exhibit non-classical signatures and can lead to interesting observation such as photon antibunching \cite{kimble1977photon}. This incoherent contribution originates from the correlations in the dipole moment. In order to access the full properties of the harmonic radiation we should not perform the approximation of neglecting the dipole moment correlations. Including those correlations one can obtain the complete properties of the light field in the process of HHG, and further allows to obtain a detailed \textit{theory of quantum optical coherence for the process of high harmonic generation.} 
Furthermore, including those correlations it allows to ask for the actual quantum state of the field after HHG, going beyond the product coherent states in \eqref{eq:HHG_final}. Taking into account terms beyond linear order in $\vb{E}_Q(t)$ would lead to a coupling of different field modes, and thus to entanglement and squeezing. \\

All the previous analysis was performed in the Schrödinger picture (or more precisely in the interaction picture). However, to compute the observables of the field, such as the spectra or two-time correlation functions, and eventually finding non-classical signatures, does not necessarily require the knowledge of the field state after the interaction.
That's why we will switch to the Heisenberg picture, making the field operators time-dependent, which allows to obtain two-time averages including the dipole moment correlations. 
We will start with the Hamiltonian of the intense-laser matter interaction (here in 1D for linear polarization)
\begin{align}
    H = \sum_q \omega_q b_q^\dagger b_q + H_A - {d} {E_Q},
\end{align}
where $H_A$ is the atomic Hamiltonian, and the electric field operator is given by ${E}_Q = - i g \sum_q \sqrt{q} (b_q^\dagger - b_q)$. First, we have to transform the field operator into the Heisenberg picture
\begin{align}
    b_q(t) = b_q e^{- i \omega_q t} + \sqrt{q} g \int_0^t dt^\prime {d}(t^\prime) e^{- i \omega_q (t-t^\prime)}.
\end{align}
We will first compute the first order correlation function \cite{carmichael1999statistical}
\begin{align}
    G(t,t+\tau) = \expval{b_q^\dagger(t) b_q(t+\tau)} = q g^2 e^{i \omega_q \tau} \int_0^t dt_1 e^{- i \omega_q t_1} \int_0^{t+\tau} dt_2 e^{i \omega_q t_2} \bra{g} {d}(t_1)  {d}(t_2) \ket{g},
\end{align}
such that we can use the Wiener-Khinchin theorem \cite{mandel1995optical}, stating that the auto-correlation function of a stationary random process and the spectral density of this process are a Fourier-transform pair in the ensemble average, to obtain the power spectrum given by 
\begin{align}
\label{eq:def_spectrum}
    S(\omega) = \frac{1}{\pi} \operatorname{Re} \left[ \int_0^\infty d \tau \lim_{t \to \infty} \expval{ b_q^\dagger(t) b_q(t+\tau)} e^{i \omega \tau} \right].
\end{align}
It turns out that the power spectral density $S(\omega)$ consists of two terms, the coherent part, and an incoherent contribution coming from the dipole moment correlations
\begin{align}
    G^{(1)} (t,t+ \tau) = & G^{(1)}_{coh} (t,t+ \tau)  + q g^2 e^{i \omega_q \tau} \int_0^t dt_1 e^{- i \omega_q t_1} \int_0^{t+\tau} dt_2 e^{i \omega_q t_2} \int dp \bra{g} d(t_1) \ket{p} \bra{p} d(t_2) \ket{g},
\end{align}
where the coherent contribution (first term) comes from the dipole moment expectation value. In the stationary limit this terms reads 
\begin{align}
    \lim_{t \to \infty} G^{(1)}_{coh} (t,t+\tau) = g^2 q \abs{\expval{{d}}(\omega_q)}^2 e^{- i \omega_q \tau },
\end{align}
such that the coherent contribution to the power spectrum is given by
\begin{align}
    S_{coh}(\omega) =  g^2 q \abs{\expval{{d}}(\omega_q)}^2 \delta(\omega - \omega_q).
\end{align}
It shows that the HHG spectrum consists of peaks at frequency $\omega_q = q \omega$ (when properly taking into account the finite duration of the driving pulse, the harmonic peaks will have a finite width), with the weight of each harmonic given by the Fourier transform of the time dependent dipole moment expectation value, and it remains to compute the incoherent contribution.
However, it also needs to be carefully analyzed if the Wiener-Khinchin theorem (WKT) can be used since it only holds for a stationary random process in the ensemble average (see discussion about time-dependent spectra in \cite{eberly1977time, eberly1980time}). One should also analyze if HHG is an ergodic process, which would then allow to use the WKT since the ensemble and time average agree for a stationary process and the autocorrelation function in \eqref{eq:def_spectrum} only depends on the temporal difference (stationarity in the ensemble or temporal average are not sufficient for ergodicity). 
Furthermore, we then want to compute the second order correlation function 
\begin{align}
    g^{(2)}(\tau) = \lim_{t \to \infty} \frac{\expval{b_q^\dagger(t) b_q^\dagger(t+\tau) b_q(t+\tau) b_q(t)}}{\expval{b_q^\dagger(t) b_q(t)} \expval{b_q^\dagger(t+\tau) b_q(t+\tau)}},
\end{align}
since this would provide insights into possible anti-bunching signatures, i.e. $g^{(2)} (0) < g^{(2)}(\tau)$. 
However, we imagine that the coherent contribution dominates the incoherent contribution, and one needs to conceive clever experiments to either separate the two processes for individual harmonics or to find the conditions in which the two contributions are on the same order of magnitude. This could eventually be realized with a two-color driving field ($\omega$ and its second harmonic $2\omega$), which leads to the appearance of even harmonics in the spectrum. By varying the phase between the two driving fields, the amplitude of the even harmonics can be altered, such that there might be a regime in which the coherent and incoherent contribution can compete.

\subsection{\label{sec:squeeze}Entanglement and squeezing in high harmonic generation}


Thus far found that the field state of the harmonic modes are given by product coherent states of all filed modes \eqref{eq:HHG_final}.
This is a consequence of the approximation performed in \eqref{eq:kraus_approx} (neglecting the dipole moment correlations), which effectively leads to a linear expression in the field operators $b_q^{(\dagger)}$. 
While the commutator of the exact interaction Hamiltonian $H_I(t) = - {d}(t) {E}_Q(t)$ at different times is an operator in the total Hilbert space of atom plus field
\begin{align}
    [H_I(t_1), H_I(t_2)] \in \mathcal{H_A} \otimes \mathcal{H}_F.    
\end{align}
The approximate interaction Hamiltonian $H_I^{app}(t) =  - \expval{d(t)} E_Q(t)$ is just a complex number, i.e. $[H_I^{app}(t_1) , H_I^{app}(t_2) ] \in \mathds{C}$, and thus when solving \eqref{eq:kraus_approx} the modes do not mix. Going beyond the linear term of the field operator ${E}_Q(t)$ would lead, for instance, to squeezing in the field modes. Furthermore, all field modes will become entangled due to the mixing of the field operators $b_q^{(\dagger)}$ of the different modes. 
We can thus start to evaluate the commutator of the exact interaction Hamiltonian at different times, yielding 
\begin{align}
\label{eq:commutator}
    [H_I(t_1), H_I(t_2)] = &  - g^2 \sum_{qp} \sqrt{qp} \sum_{ijk}  \ket{i}\bra{j}   [ d_{ik}(t_1) d_{kj}(t_2) - d_{ik}(t_2) d_{kj}(t_1) ] \left[ b_q^\dagger b_p^\dagger e^{i \omega_q t_1} e^{i \omega_p t_2} -  b_q^\dagger b_p e^{- i \omega_p t_2} e^{i \omega_q t_1} + \operatorname{h.c.}  \right] \nonumber \\
    &  + g^2 \sum_q q \sum_{ijk} [ d_{ik}(t_1) d_{kj}(t_2) e^{- i \omega_q (t_1-t_2)} - d_{ik}(t_2) d_{kj}(t_1) e^{i \omega_q (t_1-t_2)} ] \ket{i }\bra{j},
\end{align}
where we have used a discrete basis for the atomic degree of freedom $\mathds{1} = \sum_i \dyad{i}$, and introduced the transition dipole matrix elements $d_{ij}(t) = \bra{i} d(t) \ket{i}$. Note that for the approximation of neglecting the dipole moment correlations and taking the expectation value in the electronic ground state leads to $\sum_{ijk} d_{ik}(t_1) d_{kj} (t_2) \bra{g} \ket{i} \bra{j} \ket{g} \simeq \expval{d(t_1)} \expval{d(t_2)}$, and thus the first line in \eqref{eq:commutator} vanishes (where the squeezing and mixing of modes would came from), and the second line reduces to what one would get from $[H^{app}_I(t_1), H^{app}_I(t_2)].$
However, for the exact interaction Hamiltonian $H_I(t) = - {d}(t) {E}_Q(t)$, we observe that the different field modes mix, which would lead to squeezing and entanglement. 
One could, for instance, already observe first signatures of such non-classical states due to the higher order terms of ${E}_Q(t)$ when taking into up to the quadratic order in the coupling $g \propto \sqrt{\omega/V_{eff}}$ with the quantization volume $V_{eff}$. Thus, when solving \eqref{eq:kraus_exact} by using Baker-Campbell-Hausdorff for infinitesimal time steps, one obtains an approximate solution up to quadratic order in $g$ when only taking into account $[H_I(t_1), H_I(t_2)] \propto g^2$, and the time-dependent transition dipole matrix elements $d_{ij}(t)$ can be computed withing the strong field approximation \cite{amini2019symphony}.

\section{Conclusion}

Motivated by recent studies on the quantum optical description of the process of high harmonic generation from inten laser field driven atoms, we identified current challenges and how this can lead to future investigations. With the proposed studies we anticipate that more complete insights into the process of HHG will be obtained, and that the full characteristics of the radiation field are found. 
The current quantum optical framework treats the source of the scattered field as a classical charge current, similar to a dipole antenna, and thus only the coherent contribution is obtained through the dipole moment expectation value. Thus, the radiation properties as well as the final field state, do not indicate genuine quantum signatures in the HHG process. Only via conditioning experiments, through a post-selection procedure, we obtained non-classical signatures in the reconstructed Wigner function. 
It would thus be of great interest to see if already at the level of the HHG process itself, without conditioning, non-classical observations can be obtained in the radiation properties of the scattered field.  
Besides the proposed approaches present in this manuscript, there  exist further efforts in this direction. For instance there are the following options to achieve such situations:
\begin{itemize}
\item So far we have considered high-order harmonics generated in atomic systems. Alternatively, one can consider HHG from solid state targets. Even in the case of "trivial" solid state systems, such as electrons in the Wannier-Bloch picture \cite{osika2017wannier}, one can obtain electron-field entanglement \cite{rivera2022quantum} since the electron can transition on one site in the lattice, but might recombine in another side. A similar mechanism, of semiconductors driven by strong coherent radiation, is studied in the recent paper \cite{gonoskov2022nonclassical}, where the potential for generating non-classical light fields is discussed.

\item An other option, besides driving HHG in simple uncorrelated solid state targets, is to look for HHG from laser driven strongly correlated materials, such as high temperature superconductors \cite{alcala2022high}. For a simple, yet pedagogical model of such mechanism, see \cite{pizzi2023light,tzallas2023quantum}.

\item Finally, one can use non-classical, for instance squeezed light to drive the HHG process in atoms, which leads it's fingerprints in the field observable such as the HHG spectra \cite{gorlach2022high}. Which, however, also don't depict non-classical signatures in the harmonic radiation based on this observable.
\end{itemize}.

\section{Acknowledgement}

ICFO group acknowledges support from: ERC AdG NOQIA; Ministerio de Ciencia y Innovation Agencia Estatal de Investigaciones (PGC2018-097027-B-I00/10.13039/501100011033, CEX2019-000910-S/10.13039/501100011033, Plan National FIDEUA PID2019-106901GB-I00, FPI, QUANTERA MAQS PCI2019-111828-2, QUANTERA DYNAMITE PCI2022-132919, Proyectos de I+D+I “Retos Colaboración” QUSPIN RTC2019-007196-7); MICIIN with funding from European Union NextGenerationEU(PRTR-C17.I1) and by Generalitat de Catalunya; Fundació Cellex; Fundació Mir-Puig; Generalitat de Catalunya (European Social Fund FEDER and CERCA program, AGAUR Grant No. 2021 SGR 01452, QuantumCAT \ U16-011424, co-funded by ERDF Operational Program of Catalonia 2014-2020); Barcelona Supercomputing Center MareNostrum (FI-2022-1-0042); EU Horizon 2020 FET-OPEN OPTOlogic (Grant No 899794); EU Horizon Europe Program (Grant Agreement 101080086 — NeQST), National Science Centre, Poland (Symfonia Grant No. 2016/20/W/ST4/00314); ICFO Internal “QuantumGaudi” project; European Union’s Horizon 2020 research and innovation program under the Marie-Skłodowska-Curie grant agreement No 101029393 (STREDCH) and No 847648 (“La Caixa” Junior Leaders fellowships ID100010434: LCF/BQ/PI19/11690013, LCF/BQ/PI20/11760031, LCF/BQ/PR20/11770012, LCF/BQ/PR21/11840013). Views and opinions expressed in this work are, however, those of the author(s) only and do not necessarily reflect those of the European Union, European Climate, Infrastructure and Environment Executive Agency (CINEA), nor any other granting authority. Neither the European Union nor any granting authority can be held responsible for them.
P.S. acknowledges funding from The European Union’s Horizon 2020
research and innovation programme under the Marie Skłodowska-Curie grant agreement No 847517.

\bibliographystyle{unsrt}
\bibliography{literatur}

\begin{thebibliography}{10}

\bibitem{bialynicki2013quantum}
Iwo Bia{\l}ynicki-Birula and Zofia Bia{\l}ynicka-Birula.
\newblock {\em Quantum electrodynamics}, volume~70.
\newblock Elsevier, 2013.

\bibitem{zofia}
Zofia~Bia\l ynicka Birula.
\newblock Strong-field effects in electron spectra from multiphoton ionisation.
\newblock {\em Journal of Physics B: Atomic and Molecular Physics}, 17:3091,
  1984.

\bibitem{ferray1988multiple}
M~Ferray, Anne L'Huillier, XF~Li, LA~Lompre, G~Mainfray, and C~Manus.
\newblock Multiple-harmonic conversion of 1064 nm radiation in rare gases.
\newblock {\em Journal of Physics B: Atomic, Molecular and Optical Physics},
  21(3):L31, 1988.

\bibitem{lewenstein1994theory}
Maciej Lewenstein, Ph~Balcou, M~Yu Ivanov, Anne L’huillier, and Paul~B
  Corkum.
\newblock Theory of high-harmonic generation by low-frequency laser fields.
\newblock {\em Physical Review A}, 49(3):2117, 1994.

\bibitem{agostini1979free}
Pierre Agostini, F~Fabre, G{\'e}rard Mainfray, Guillaume Petite, and N~Ko
  Rahman.
\newblock Free-free transitions following six-photon ionization of xenon atoms.
\newblock {\em Physical Review Letters}, 42(17):1127, 1979.

\bibitem{lewenstein1995rings}
M~Lewenstein, KC~Kulander, KJ~Schafer, and PH~Bucksbaum.
\newblock Rings in above-threshold ionization: A quasiclassical analysis.
\newblock {\em Physical Review A}, 51(2):1495, 1995.

\bibitem{corkum1993plasma}
Paul~B Corkum.
\newblock Plasma perspective on strong field multiphoton ionization.
\newblock {\em Physical review letters}, 71(13):1994, 1993.

\bibitem{salieres2001feynman}
Pascal Sali{\`e}res, B~Carr{\'e}, L~Le~D{\'e}roff, F~Grasbon, GG~Paulus,
  H~Walther, R~Kopold, W~Becker, DB~Milosevic, A~Sanpera, et~al.
\newblock Feynman's path-integral approach for intense-laser-atom interactions.
\newblock {\em Science}, 292(5518):902--905, 2001.

\bibitem{ivanov2005anatomy}
Misha~Yu Ivanov, Michael Spanner, and Olga Smirnova.
\newblock Anatomy of strong field ionization.
\newblock {\em Journal of Modern Optics}, 52(2-3):165--184, 2005.

\bibitem{smirnova2013multielectron}
Olga Smirnova and Misha Ivanov.
\newblock Multielectron high harmonic generation: simple man on a complex
  plane.
\newblock {\em arXiv preprint arXiv:1304.2413}, 2013.

\bibitem{amini2019symphony}
Kasra Amini, Jens Biegert, Francesca Calegari, Alexis Chac{\'o}n, Marcelo~F
  Ciappina, Alexandre Dauphin, Dmitry~K Efimov, Carla~Figueira
  de~Morisson~Faria, Krzysztof Giergiel, Piotr Gniewek, et~al.
\newblock Symphony on strong field approximation.
\newblock {\em Reports on Progress in Physics}, 82(11):116001, 2019.

\bibitem{gorlach2020quantum}
Alexey Gorlach, Ofer Neufeld, Nicholas Rivera, Oren Cohen, and Ido Kaminer.
\newblock The quantum-optical nature of high harmonic generation.
\newblock {\em Nature communications}, 11(1):1--11, 2020.

\bibitem{lewenstein2021generation}
M~Lewenstein, MF~Ciappina, E~Pisanty, J~Rivera-Dean, P~Stammer, Th~Lamprou, and
  P~Tzallas.
\newblock Generation of optical schr{\"o}dinger cat states in intense
  laser--matter interactions.
\newblock {\em Nature Physics}, 17(10):1104--1108, 2021.

\bibitem{rivera2022strong}
J~Rivera-Dean, Th~Lamprou, E~Pisanty, P~Stammer, AF~Ord{\'o}{\~n}ez,
  AS~Maxwell, MF~Ciappina, M~Lewenstein, and P~Tzallas.
\newblock Strong laser fields and their power to generate controllable
  high-photon-number coherent-state superpositions.
\newblock {\em Physical Review A}, 105(3):033714, 2022.

\bibitem{gonoskov2016quantum}
IA~Gonoskov, N~Tsatrafyllis, IK~Kominis, and P~Tzallas.
\newblock Quantum optical signatures in strong-field laser physics: Infrared
  photon counting in high-order-harmonic generation.
\newblock {\em Scientific Reports}, 6(1):1--9, 2016.

\bibitem{tsatrafyllis2017high}
N~Tsatrafyllis, IK~Kominis, IA~Gonoskov, and P~Tzallas.
\newblock High-order harmonics measured by the photon statistics of the
  infrared driving-field exiting the atomic medium.
\newblock {\em Nature communications}, 8(1):1--6, 2017.

\bibitem{stammer2022high}
Philipp Stammer, Javier Rivera-Dean, Theocharis Lamprou, Emilio Pisanty,
  Marcelo~F Ciappina, Paraskevas Tzallas, and Maciej Lewenstein.
\newblock High photon number entangled states and coherent state superposition
  from the extreme ultraviolet to the far infrared.
\newblock {\em Physical Review Letters}, 128(12):123603, 2022.

\bibitem{stammer2022theory}
Philipp Stammer.
\newblock Theory of entanglement and measurement in high-order harmonic
  generation.
\newblock {\em Physical Review A}, 106(5):L050402, 2022.

\bibitem{stammer2022quantum}
Philipp Stammer, Javier Rivera-Dean, Andrew Maxwell, Theocharis Lamprou,
  Andr{\'e}s Ord{\'o}{\~n}ez, Marcelo~F Ciappina, Paraskevas Tzallas, and
  Maciej Lewenstein.
\newblock Quantum electrodynamics of intense laser-matter interactions: A tool
  for quantum state engineering.
\newblock {\em PRX Quantum}, 4(1):010201, 2023.

\bibitem{rivera2022light}
Javier Rivera-Dean, Philipp Stammer, Andrew~S Maxwell, Th~Lamprou, Paraskevas
  Tzallas, Maciej Lewenstein, and Marcelo~F Ciappina.
\newblock Light-matter entanglement after above-threshold ionization processes
  in atoms.
\newblock {\em Physical Review A}, 106(6):063705, 2022.

\bibitem{sundaram1990high}
Bala Sundaram and Peter~W Milonni.
\newblock High-order harmonic generation: simplified model and relevance of
  single-atom theories to experiment.
\newblock {\em Physical Review A}, 41(11):6571, 1990.

\bibitem{lewenstein2022attosecond}
Maciej Lewenstein, Niccolo Baldelli, Utso Bhattacharya, Jens Biegert,
  Marcelo~Fabi{\'a}n Ciappina, Ugaitz Elu, T~Grass, PT~Grochowski, A~Johnson,
  Th~Lamprou, et~al.
\newblock Attosecond physics and quantum information science.
\newblock {\em arXiv preprint arXiv:2208.14769}, 2022.

\bibitem{bhattacharya2023strong}
Utso Bhattacharya, Theocharis Lamprou, Andrew~S Maxwell, Andr{\'e}s~F
  Ord{\'o}{\~n}ez, Emilio Pisanty, Javier Rivera-Dean, Philipp Stammer,
  Marcelo~F Ciappina, Maciej Lewenstein, and Paraskevas Tzallas.
\newblock Strong laser physics, non-classical light states and quantum
  information science.
\newblock {\em arXiv preprint arXiv:2302.04692}, 2023.

\bibitem{salieres1997study}
Pascal Salieres, Anne L'huillier, Philippe Antoine, and Maciej Lewenstein.
\newblock Study of the spatial and temporal coherence of high order harmonics.
\newblock {\em arXiv preprint quant-ph/9710060}, 1997.

\bibitem{eberly1992spectrum}
JH~Eberly and MV~Fedorov.
\newblock Spectrum of light scattered coherently or incoherently by a
  collection of atoms.
\newblock {\em Physical Review A}, 45(7):4706, 1992.

\bibitem{diestler2008harmonic}
DJ~Diestler.
\newblock Harmonic generation: quantum-electrodynamical theory of the harmonic
  photon-number spectrum.
\newblock {\em Physical Review A}, 78(3):033814, 2008.

\bibitem{kominis2014quantum}
IK~Kominis, G~Kolliopoulos, D~Charalambidis, and P~Tzallas.
\newblock Quantum-optical nature of the recollision process in
  high-order-harmonic generation.
\newblock {\em Physical Review A}, 89(6):063827, 2014.

\bibitem{krausz2009attosecond}
Ferenc Krausz and Misha Ivanov.
\newblock Attosecond physics.
\newblock {\em Reviews of modern physics}, 81(1):163, 2009.

\bibitem{li2022experimental}
Cheng Li, Boris Braverman, Girish Kulkarni, and Robert~W Boyd.
\newblock Experimental generation of polarization entanglement from spontaneous
  parametric down-conversion pumped by spatiotemporally highly incoherent
  light.
\newblock {\em arXiv preprint arXiv:2210.16229}, 2022.

\bibitem{molmer1997optical}
Klaus M{\o}lmer.
\newblock Optical coherence: A convenient fiction.
\newblock {\em Physical Review A}, 55(4):3195, 1997.

\bibitem{van2001quantum}
SJ~Van~Enk and Christopher~A Fuchs.
\newblock Quantum state of an ideal propagating laser field.
\newblock {\em Physical review letters}, 88(2):027902, 2001.

\bibitem{van2001quantum2}
Steven~J van Enk and Christopher~A Fuchs.
\newblock The quantum state of a propagating laser field.
\newblock {\em arXiv preprint quant-ph/0111157}, 2001.

\bibitem{bartlett2006dialogue}
Stephen~D Bartlett, Terry Rudolph, and Robert~W Spekkens.
\newblock Dialogue concerning two views on quantum coherence: factist and
  fictionist.
\newblock {\em International Journal of Quantum Information}, 4(01):17--43,
  2006.

\bibitem{carmichael1999statistical}
Howard Carmichael.
\newblock {\em Statistical methods in quantum optics 1: master equations and
  Fokker-Planck equations}, volume~1.
\newblock Springer Science \& Business Media, 1999.

\bibitem{kimble1977photon}
H~Jeff Kimble, Mario Dagenais, and Leonard Mandel.
\newblock Photon antibunching in resonance fluorescence.
\newblock {\em Physical Review Letters}, 39(11):691, 1977.

\bibitem{mandel1995optical}
Leonard Mandel and Emil Wolf.
\newblock {\em Optical coherence and quantum optics}.
\newblock Cambridge university press, 1995.

\bibitem{eberly1977time}
JH~Eberly and K~Wodkiewicz.
\newblock The time-dependent physical spectrum of light.
\newblock {\em JOSA}, 67(9):1252--1261, 1977.

\bibitem{eberly1980time}
JH~Eberly, CV~Kunasz, and K~Wodkiewicz.
\newblock Time-dependent spectrum of resonance fluorescence.
\newblock {\em Journal of Physics B: Atomic and Molecular Physics (1968-1987)},
  13(2):217, 1980.

\bibitem{osika2017wannier}
Edyta~N Osika, Alexis Chac{\'o}n, Lisa Ortmann, Noslen Su{\'a}rez, Jose~Antonio
  P{\'e}rez-Hern{\'a}ndez, Bart{\l}omiej Szafran, Marcelo~F Ciappina, Fernando
  Sols, Alexandra~S Landsman, and Maciej Lewenstein.
\newblock Wannier-bloch approach to localization in high-harmonics generation
  in solids.
\newblock {\em Physical Review X}, 7(2):021017, 2017.

\bibitem{rivera2022quantum}
Javier Rivera-Dean, Philipp Stammer, Andrew~S Maxwell, Theocharis Lamprou,
  Andr{\'e}s~F Ord{\'o}{\~n}ez, Emilio Pisanty, Paraskevas Tzallas, Maciej
  Lewenstein, and Marcelo~F Ciappina.
\newblock Quantum optical analysis of high-harmonic generation in solids within
  a wannier-bloch picture.
\newblock {\em arXiv preprint arXiv:2211.00033}, 2022.

\bibitem{gonoskov2022nonclassical}
Ivan Gonoskov, Ren{\'e} Sondenheimer, Christian H{\"u}necke, Daniil Kartashov,
  Ulf Peschel, and Stefanie Gr{\"a}fe.
\newblock Nonclassical light generation and control from laser-driven
  semiconductor intraband excitations.
\newblock {\em arXiv preprint arXiv:2211.06177}, 2022.

\bibitem{alcala2022high}
Jordi Alcal{\`a}, Utso Bhattacharya, Jens Biegert, Marcelo Ciappina, Ugaitz
  Elu, Tobias Gra{\ss}, Piotr~T Grochowski, Maciej Lewenstein, Anna Palau,
  Themistoklis~PH Sidiropoulos, et~al.
\newblock High-harmonic spectroscopy of quantum phase transitions in a high-tc
  superconductor.
\newblock {\em Proceedings of the National Academy of Sciences},
  119(40):e2207766119, 2022.

\bibitem{pizzi2023light}
Andrea Pizzi, Alexey Gorlach, Nicholas Rivera, Andreas Nunnenkamp, and Ido
  Kaminer.
\newblock Light emission from strongly driven many-body systems.
\newblock {\em Nature Physics}, pages 1--11, 2023.

\bibitem{tzallas2023quantum}
Paraskevas Tzallas.
\newblock Quantum correlated atoms in intense laser fields.
\newblock {\em Nature Physics}, pages 1--2, 2023.

\bibitem{gorlach2022high}
Alexey Gorlach, Matan~Even Tzur, Michael Birk, Michael Kr{\"u}ger, Nicholas
  Rivera, Oren Cohen, and Ido Kaminer.
\newblock High harmonic generation driven by quantum light.
\newblock {\em arXiv preprint arXiv:2211.03188}, 2022.

\end{thebibliography}

\end{document}